\documentclass[amsmath,amssymb,nofootinbib]{revtex4}
\usepackage{graphicx}
\usepackage{epsfig}
\begin{document}
\title{Time in quantum theory, the Wheeler-DeWitt equation and  the Born-Oppenheimer approximation}
\author{Alexander Yu. Kamenshchik}
\email{kamenshchik@bo.infn.it}
\affiliation{Dipartimento di Fisica e Astronomia, Universit\`a di Bologna and INFN, via Irnerio 46, 40126 Bologna, Italy}
\affiliation{L.D. Landau Institute for Theoretical Physics of the Russian
Academy of Sciences, Kosygin str. 2, 119334 Moscow, Russia}
\author{Alessandro Tronconi}
\email{tronconi@bo.infn.it}
\affiliation{Dipartimento di Fisica e Astronomia, Universit\`a di Bologna and INFN, via Irnerio 46, 40126 Bologna, Italy}
\author{Tereza Vardanyan}
\email{tereza.vardanyan@bo.infn.it}
\affiliation{Dipartimento di Fisica e Astronomia, Universit\`a di Bologna and INFN, via Irnerio 46, 40126 Bologna, Italy}
\author{Giovanni Venturi}
\email{giovanni.venturi@bo.infn.it}
\affiliation{Dipartimento di Fisica e Astronomia, Universit\`a di Bologna and INFN, via Irnerio 46, 40126 Bologna, Italy}

\begin{abstract}
We compare two different approaches to the treatment of the Wheeler-DeWitt equation and the introduction of time in quantum cosmology.
One approach is based on the gauge-fixing procedure in  theories with  first-class constraints, while the other uses the Born-Oppenheimer method.
We apply both to a very simple cosmological model and observe that they give similar predictions. We also discuss the problem of time in non-relativistic quantum mechanics and some questions concerning the correspondence between classical and quantum theories. 
\end{abstract}

\maketitle

\section{Introduction}
The problem of ``disappearance of time'' in quantum gravity and cosmology is well known and has a rather  long history (see e.g. \cite{Kuchar, Kiefer} and references therein).
Let us remember briefly what it means. The Hilbert-Einstein Lagrangian of General Relativity contains the Lagrange multipliers $N$ and $N^i$, which are called  lapse and  shift functions. There are also the dynamical degrees of freedom connected with the spatial components of the metric $g_{ij}$ and with the non-gravitational fields present in the universe. One introduces the conjugate momenta and makes a Legendre transformation in order to use the canonical formalism. Then one discovers that the Hamiltonian is proportional to the linear combination of the constraints, multiplied by the Lagrange multipliers \cite{DeWitt}.  Thus, the Hamiltonian vanishes if the constraints are satisfied. This can be interpreted as the impossibility of writing  down a time-dependent Schr\"odinger equation. One can see this problem from a somewhat different point of view. If one applies 
the Dirac quantisation procedure, then the constraint, wherein the classical phase variables are substituted by the quantum operators, should annihilate the quantum state of the system under consideration.  Gravitational constraints contain  momenta, which classically are  time derivatives of  fields, but their origin connected with the classical notion of time vanishes in the quantum theory, where momenta are simply operators satisfying some commutation relations. The main constraint arising in General Relativity 
is quadratic in momenta and gives rise to the so called Wheeler-DeWitt equation \cite{DeWitt}. It is time-independent, but we know that the universe lives in time. Where is it hidden? We shall consider two approaches to this problem and apply them to a very simple toy model in order to understand how  they are related. 

The first approach is based on the fact that when we have   a system with first class constraints it is necessary to add  some gauge-fixing conditions to them (see e.g. \cite{Sundermeyer, Henneaux}).  If we choose some time-dependent classical gauge-fixing condition,  we can define a classical time parameter, expressed by means of a certain combination of phase variables. Some other variables are excluded from the game by resolving the constraints. As a result one obtains a non-vanishing  effective physical Hamiltonian, which depends on the remaining phase variables which are quantised. This Hamiltonian governs the evolution of the wave function, describing the degrees of freedom, which did not participate in the definition of time. Then one can reconstruct a particular  solution of the initial Wheeler-DeWitt equation, corresponding to the solution of the physical Schr\"odinger equation. The above approach was elaborated in detail in  paper \cite{Barv}. In  paper \cite{Bar-Kam}!
  it was applied to some very simple 
cosmological models explicitly. It is important to note that in this approach the time and the Hamiltonian are introduced before the quantisation of the physical degrees of freedom. 

The second approach is based on the Born-Oppenheimer method \cite{Brout, Brout-Venturi,Venturi}, which was very effective in the treatment of atoms and molecules \cite{BO}.   
This method relies on the existence of two time  or energy scales. In the case of cosmology one of these scales is related to the Planck mass and to a slow 
evolution of gravitational degrees of freedom. The other one is characterised by a much smaller energy and by a fast motion of matter degrees of freedom. 
The solution to the Wheeler-DeWitt equation is factorised into a product of two wave functions. One of them  has a semiclassical structure, where the action satisfies some kind of  Einstein-Hamilton-Jacobi equation and the average of the quantum matter energy-momentum tensor serves as a source for  gravity. The semiclassical time arises from this equation. The second wave function describes  quantum degrees of freedom of  matter and satisfies the effective Schr\"odinger equation. 
The  latter  are responsible for the introduction of the time parameter.   However, the principles of  separation between these two kinds of the degrees of freedom are different. Thus, it would be interesting to compare these two methods by applying them to a simple toy model. We shall do exactly that in this short note. Its second section  is devoted to the study of  a  simple cosmological model.
In the third section we discuss the problem of time in non-relativistic quantum mechanics and some aspects of the correspondence between classical and quantum theories.
The last section contains  concluding remarks.  

\section{A very simple toy model}
\subsection{Classical dynamics}
Let us consider the simplest toy model -- a flat Friedmann universe with the metric 
\begin{equation}
ds^2=N^2(t)dt^2 - a^2(t)dl^2,
\label{Fried}
\end{equation} 
where $N$ is, as usual, the lapse function and $a(t)$ is the scale factor.
This universe is filled with  a massless spatially homogeneous scalar field $\phi(t)$ minimally coupled to gravity. 
For this minisuperspace model the Lagrangian can be written as 
\begin{equation}
 {\cal L} = -\frac{L^3M^2\dot{a}^2a}{2N}+\frac{L^3\dot{\phi}^2 a^3}{2N},
 \label{Lagrange}
 \end{equation}
 where $M$ is a conveniently rescaled Planck mass, and $L$ is the length scale. 
 It will be convenient to use another parametrisation of the scale factor 
 \begin{equation}
 a(t) = e^{\alpha(t)}.
 \label{scale}
 \end{equation}
 Then 
 \begin{equation}
 {\cal L} = -\frac{L^3M^2\dot{\alpha}^2e^{3\alpha}}{2N}+\frac{L^3\dot{\phi}^2 e^{3\alpha}}{2N}.
 \label{Lagrange1}
 \end{equation} 
 The variation of the Lagrangian (\ref{Lagrange1}) with respect to the lapse function $N$ gives the first Friedmann equation 
 \begin{equation}
 M^2\dot{\alpha^2} = \dot{\phi}^2,
 \label{Fried1}
 \end{equation}
 while its  variation with respect to $\phi$ gives the first integral of the Klein-Gordon equation 
 \begin{equation}
 \frac{L^3\dot{\phi}e^{3\alpha}}{N} = p_{\phi} = const.
 \label{KG}
 \end{equation}
 Here, $p_{\phi}$ is the conjugate momentum, which is conserved during the classical time evolution of our universe.
 It will be convenient to choose as a time parameter $t$, the cosmic time, which is equivalent to  fixing $N = 1$.
 Substituting Eq. (\ref{KG}) into Eq. (\ref{Fried1}), we find for the expanding universe and $0 \leq t \leq \infty$ 
 \begin{equation}
 e^{3\alpha} = 3\frac{|p_{\phi}|}{ML^3}t,
 \label{expand}
 \end{equation}
 and for the contracting universe, when $-\infty < t \leq 0$:
 \begin{equation}
 e^{3\alpha} = -3\frac{|p_{\phi}|}{ML^3}t.
 \label{contract}
 \end{equation}
 In what follows we shall consider the expanding universe and we shall choose the positive sign for $p_{\phi}$ without losing generality. 
 
 \subsection{Wheeler-DeWitt equation and the gauge fixing procedure}
 On introducing the conjugate momenta $p_{\phi}$ (see Eq. (\ref{KG})) and 
 \begin{equation}
 p_{\alpha} = -\frac{L^3M^2\dot{\alpha}e^{3\alpha}}{N},
 \label{momentum}
 \end{equation}
 and making the Legendre transformation, we see that the Hamiltonian is 
 \begin{equation}
 {\cal H} = N\left(-\frac{p_{\alpha}^2e^{-3\alpha}}{2M^2}+\frac{p_{\phi}^2e^{-
 3\alpha}}{2}\right) = NH,
 \label{Hamilton}
 \end{equation}
 where $H$ is the so called super-Hamiltonian constraint. From equations (\ref{Fried1}), (\ref{KG}) and (\ref{momentum}) it is obvious that the Hamiltonian is constrained to vanish:
 \begin{equation}
 H=0.
 \label{constr}
 \end{equation} 
 The action in the Hamiltonian form is 
 \begin{equation}
 S = \int dt (p_{\alpha}\dot{\alpha}+p_{\phi}\dot{\phi}-NH).
 \label{action-Ham}
 \end{equation} 
 
 On performing the procedure of the Dirac quantisation of the system with constraints \cite{Dirac}, we obtain the Wheeler-DeWitt equation:
 \begin{equation}
 \hat{H}|\Psi\rangle = 0.  
 \label{WdW}
 \end{equation}
 Here, the operator $\hat{H}$ arises when we substitute  the phase variables by the corresponding operators and fix some particular operator ordering.
 $|\Psi\rangle$ is a quantum state of the Universe.
 Now, we shall choose the simplest operator ordering, such that 
 \begin{equation}
 \hat{H} = e^{-3\hat{\alpha}}\left(-\frac{\hat{p}_{\alpha}^2}{2M^2}+\frac{\hat{p}_{\phi}^2}{2}\right).
 \label{WdW1}
 \end{equation}
 Further, it will be convenient to  consider the quantum state $|\Psi\rangle$ in the $(\alpha,p_{\phi})$ representation. Thus,
 the Wheeler-DeWitt equation shall have the following form:
 \begin{equation}
 \left( \frac{\partial^2}{\partial\alpha^2}+M^2p_{\phi}^2\right)\Psi(\alpha,p_{\phi})=0.
 \label{WdW2}
 \end{equation}
 The general solution of this equation is 
 \begin{equation}
 \Psi(\alpha,p_{\phi}) = \psi_1(p_{\phi_1})e^{iM|p_{\phi}|\alpha}+\psi_1(p_{\phi_2})e^{-iM|p_{\phi}|\alpha}.
 \label{WdW-sol}
 \end{equation}
 
 We shall now obtain the effective Schr\"odiger equation for the physical wave function and the physical Hamiltonian  
  following the recipe, described in detail in \cite{Barv,Bar-Kam}. First of all we have to introduce a time-dependent gauge-fixing condition.
  Let us try to use the following one:
 \begin{equation}
 \xi(\alpha,p_{\alpha},t) = \frac{L^3M^2e^{3\alpha}}{3p_{\alpha}}-t = 0.
 \label{gauge}
 \end{equation}
 This gauge condition coincides with the classical solution of the Friedmann equation, giving the dependence of the scale factor $a$ on the cosmic time $t$.
 Then, on requiring  the conservation of the gauge condition in time and using the equation
 \begin{equation}
 \frac{d\xi}{dt}=\frac{\partial \xi}{\partial t} + N\{\xi,H\} = 0,
 \label{lapse}
 \end{equation}
 where the curly braces mean the Poisson brackets, one can easily see that the lapse function is equal to one as it should be. 
If we solve the constraint (\ref{constr}) and use the gauge-fixing condition (\ref{gauge}) to define the time, the action (\ref{action-Ham}) can be written in the following form
\begin{equation}
 S = \int dt (p_{\phi}\dot{\phi}-H_{\rm phys}),
 \label{action1-Ham}
 \end{equation} 
where the physical Hamiltonian is
 \begin{equation}
 H_{\rm phys} = \frac{Mp_{\phi}}{3t}.
 \label{Hamilton2}
 \end{equation}
The corresponding Schr\"odiger equation is 
\begin{equation}
i\frac{\partial\psi_{\rm phys}(p_{\phi},t)}{\partial t} = H_{\rm phys}\psi_{\rm phys}(p_{\phi},t) = \frac{Mp_{\phi}}{3t}\psi_{\rm phys}(p_{\phi},t).
\label{Schrod}
\end{equation}
The solution of this equation is 
\begin{equation}
\psi_{\rm phys}(p_{\phi},t) = \tilde{\psi}(p_{\phi})e^{-\frac{iMp_{\phi}}{3}\ln t}.
\label{Schrod1}
\end{equation}
Using the equation (\ref{gauge}) one can express the time $t$ as a function of the variables $\alpha$ and $p_{\phi}$ and we shall come back to one of the two branches of the general solution of the Wheeler-DeWitt equation (\ref{WdW-sol}). 
Let us note that the probability density $\tilde{\psi}^*\tilde{\psi}$, corresponding to the function (\ref{Schrod1}) does not depend on the cosmic time $t$ and this function can be normalized.  

Interestingly, we can take one of the branches of the general solution of the Wheeler-DeWitt equation and express the variable $\alpha$ as a function of time and of the variable 
$p_{\phi}$. On considering this function as the physical wave function, satisfying the Schr\"odinger equation, we can calculate its partial time derivative to find 
the physical Hamiltonian. The results will coincide with those of Eq. (\ref{Hamilton2}). 

\subsection{Born-Oppenheimer approach}
Let us suppose that in our model the solution of the Wheeler-DeWitt equation could be represented in the form \cite{Brout,Brout-Venturi} :
\begin{equation}
\Psi(p_{\phi},\alpha) = \varphi(\alpha)\chi(p_{\phi},\alpha),
\label{BO}
\end{equation}
where the function $\chi$ is considered as the wave function depending on the quantum variable $p_{\phi}$ and on the classical variable $\alpha$.
 The function $\chi$ is normalisable with respect to the    
 variable $p_{\phi}$. On substituting the decomposition (\ref{BO}) into the Wheeler-DeWitt equation (\ref{WdW2}), one obtains
 \begin{equation}
 \frac{\partial^2\varphi}{\partial\alpha^2}\chi+2\frac{\partial\varphi}{\partial\alpha}\frac{\partial\chi}{\partial\alpha}+\varphi\frac{\partial^2\chi}{\partial\alpha^2}+M^2p_{\phi}^2\varphi\chi=0.
 \label{BO1}
 \end{equation}
 Let us now suppose that the term $\varphi\frac{\partial^2\chi}{\partial\alpha^2}$ in the equation above is small with respect to other terms and omit it. 
 Further, let us suppose that the average
 \begin{equation}
 \left\langle \chi\Big|\frac{\partial \chi}{\partial \alpha}\right\rangle=0.
 \label{average}
 \end{equation}
 (As a matter of fact it is not necessary to impose the condition (\ref{average}). One can simply redefine the wave function, using the notion of the geometric phase and this condition will be satisfied authomatically \cite{Brout-Venturi, Venturi}. ) 
 Then, on multiplying Eq. (\ref{BO1}) by $\chi^*$ and averaging with respect to $p_{\phi}$ at any fixed value of $\alpha$, we obtain
 \begin{equation}
 \frac{\partial^2\varphi}{\partial\alpha^2}+M^2\langle p_{\phi}^2\rangle\varphi = 0.
 \label{BO2}
 \end{equation}
  Let us choose as a solution of this equation 
  \begin{equation}
  \varphi(\alpha) = \exp\left(-iM\sqrt{\langle p_{\phi}^2\rangle}\alpha\right).
  \label{varphi}
  \end{equation}
 On substituting the expression (\ref{varphi}) into Eq. (\ref{BO1}) , (where we have neglected the term $\varphi\frac{\partial^2\chi}{\partial\alpha^2}$), we obtain
 \begin{equation}
 -M^2\langle p_{\phi}^2\rangle \chi - 2iM\sqrt{\langle p_{\phi}^2\rangle}\frac{\partial\chi}{\partial\alpha}+M^2p_{\phi}^2\chi = 0.
 \label{BO3}
 \end{equation} 
 Coming back now to the classical Friedmann equation (classical limit of Eq. (\ref{BO2})) and substituting its classical right-hand side representing matter by the quantum average of the corresponding operator, we obtain
 \begin{equation}
 \dot{\alpha}=\frac{\sqrt{\langle p_{\phi}^2\rangle}e^{-3\alpha}}{ML^3}.
 \label{Fried-semi}
 \end{equation}
 We can say that in this way we have introduced a semiclassical cosmic time parameter. 
Using Eq. (\ref{Fried-semi}) we can rewrite Eq. (\ref{BO3}) in the following manner:
\begin{equation}
M^2\left(-\langle p_{\phi}^2\rangle\chi-2iL^3e^{3\alpha}\dot{\alpha}\frac{\partial\chi}{\partial\alpha}+p_{\phi}^2\chi\right)=0.
\label{BO4}
\end{equation} 
The presence of the factor $M^2$ in all three terms in Eq. (\ref{BO4}) shows why we could neglect 
 the term $\varphi\frac{\partial^2\chi}{\partial\alpha^2}$ in Eq. (\ref{BO1}). On now, using the equality 
 \begin{equation}
 \dot{\alpha}\frac{\partial\chi}{\partial\alpha} \equiv \frac{\partial\chi}{\partial t},
 \label{time}
 \end{equation}
 we can rewrite Eq. (\ref{BO4}) in a ``Schr\"odinger-like'' form
 \begin{equation}
 i\frac{\partial\chi}{\partial t}=H_{\phi}\chi-\langle H_{\phi}\rangle\chi,
 \label{time1}
 \end{equation}  
where the scalar field Hamiltonian $H_{\phi}$ is quadratic in the momentum as usual:
\begin{equation}
H_{\phi}=\frac{p_{\phi}^2e^{-3\alpha}}{2L^3}.
\label{Hamilton3}
\end{equation}
On now introducing 
\begin{equation}
\chi = \tilde{\chi}e^{i\int\langle H_{\phi}\rangle dt}, 
\label{chi-new}
\end{equation}
one sees that the new wave function $\tilde{\chi}$ satisfies the Schr\"odinger equation
 \begin{equation}
 i\frac{\partial\tilde{\chi}}{\partial t}=H_{\phi}\tilde{\chi}.
 \label{time1}
 \end{equation}  
 
 The variable $\alpha$ as a function of time is  found from the semiclassical Friedmann equation,  
where the classical energy density  is replaced by  the average of the quantum Hamiltonian of matter  with respect to the corresponding quantum state.
 Thus,
 \begin{equation}
 e^{3\alpha} = \frac{3\sqrt{\langle p_{\phi}^2\rangle}t}{ML^3}.
 \label{Fried-quant}
 \end{equation}
 On substituting the expression (\ref{Fried-quant}) into the Hamiltonian (\ref{Hamilton3}) we obtain
 \begin{equation}
 H_{\phi}=\frac{Mp_{\phi}^2}{3\sqrt{\langle p_{\phi}^2\rangle}t}.
 \label{Hamilton4}
 \end{equation}
 The general solution of  Eq. (\ref{time1}) with the Hamiltonian (\ref{Hamilton4}) is 
 \begin{equation}
 \chi(p_{\phi},t) = \chi_1(p_{\phi})\exp\left(-i\frac{Mp_{\phi}^2}{3\sqrt{\langle p_{\phi}^2\rangle}}\ln t\right).      
\label{solution-BO}
\end{equation}   
One can see that the structure of the solution (\ref{solution-BO}) is similar to that of the solution (\ref{Schrod1}). The corresponding probability densities coincide if  
the time-independent functions $\chi_1(p_{\phi})$ and  $\tilde{\psi}(p_{\phi})$ coincide.  

\section{The problem of time in  non-relativistic quantum mechanics and the quantum-classical correspondence}
The problem of time, quite similar to that discussed in the connection with quantum cosmology and the Wheeler-DeWitt equation, can be already studied   in 
non-relativistic quantum mechanics. Let us imagine an isolated quantum system, which finds itself in an energy eigenstate. Then   its wave function is 
\begin{equation}
\Psi(x_A,t) = e^{-iEt}\psi(x_A),
\label{non-rel}
\end{equation}
where the time parameter appears only in the phase factor, which does not depend on the variables $x_A$ and is not essential for the definition of the
quantum state. As is well-know  quantum states are determined up to a constant complex phase.  In other words all the probability distributions are independent of time.
This situation  just coincides with that of the Wheeler-DeWitt  equation with the peculiarity that in the case of the Wheeler-DeWitt equation the value of $E$ is always equal to zero. The problem of time in quantum mechanics and its analogy with the absence of time in the Wheeler-DeWitt equation was analysed in some detail in paper \cite{Englert}. 
If the set of variables includes more than one element, we can introduce an effective time parameter, identifying it with a certain function of the variables $x_A$ (a quantum clock). This topic now attracts  much attention by researchers working in  areas  which are at first glance rather distant from quantum gravity and cosmology. 
In a recent preprint \cite{Schild} a rather detailed review devoted to the appearance of time in quantum mechanics is presented.  The main idea is the following: let us consider a quantum system consisting of two subsystems, whose wave function satisfies a time-independent Schr\"odinger equation with a fixed value of  energy. One can always represent the wave function as a product of two functions: 
\begin{equation}
\psi(R,r) = \chi(R)\phi(r|R).
\label{exact-factor}
\end{equation}
Here the function $\chi(R)$ describes a subsystem, which plays the role of  ``quantum clock'', while $\phi(r|R)$ describes the subsystem, whose evolution is traced by the quantum clock. The expression $|\chi(R)|^2$ gives the marginal probability density for the quantum clock and   $|\phi(r|R)|^2$ gives the conditional probability for 
the system under consideration \cite{Hunter, factor}. Then, performing the operations which are usually connected with the Born-Oppenheimer approach, one can obtain 
the so called clock-dependent Schr\"odinger equation \cite{Schild} for the subsystem under consideration, having the form 
\begin{equation}
A\frac{d\ln \chi}{dR}\frac{\partial \phi}{\partial R} = H_{\rm eff}\phi.
\label{clock}
\end{equation}
Here, we have chosen a convenient gauge fixing of the phase in the decomposition (\ref{exact-factor}), $A$ is some coefficient and $H_{\rm eff}$ is an effective Hamiltonian for this subsystem. Let us note, that this equation is more general than the effective time-dependent Schr\"odinger equation arising in the Born-Oppenheimer approach to  molecular physics or to  cosmology and the left-hand side of Eq. (\ref{clock}) does not represent a partial derivative with respect to a time parameter. However, if the clock   has some particular semiclassical properties and if the wave function $\chi$ has a semiclassical form $\chi \sim \exp(iS)$, where $S$ is a classical action, then the left-hand side of Eq. (\ref{clock}) behaves as a partial derivative with respect to the classical time. In the approach reviewed in the paper \cite{Schild} one can underline two important features:
first, the exact factorisation of the wave function (\ref{exact-factor}) is always possible and it is always possible to obtain the clock-dependent Schr\"odinger equation from the time-independent Schr\"odinger equation. Second, a quantum clock does not always give rise to (semi)-classical time. 
In some situations it is necessary to use a coarse-graining procedure to obtain a (semi)-classical time from a quantum clock. 
 In cosmology the corresponding  models were studied in the papers 
\cite{TVV, we-previous}.   

We  would also like  to note some important difference between the transition from the time-independent Schr\"odinger equation to the clock-dependent (and then to 
time-dependent) Schr\"odinger equation in quantum mechanics (see \cite{Schild} and references therein) and the transition from the Wheeler-DeWitt equation to the time-dependent Schr\"odinger equation in quantum cosmology. Indeed,  the wave function $\psi(R,r)$ in Eq. (\ref{exact-factor}) is normalizable and both the system under consideration and the clock can be treated  to some extent on equal footing. For the case of the Wheeler-DeWitt equation, its solutions are non-renormalizable, because 
the configuration space on which they are defined contains some superfluous (gauge or non-physical) degrees of freedom \cite{Barv, Bar-Kam}. Thus, in both approaches the variables present in the Wheeler-DeWitt equation are not treated on equal footing.   
Using  the notations of Eq. (\ref{exact-factor}), we can say that the variables $r$  should be chosen to make the wave function $\phi(r|R)$ normalisable and to satisfy the Schr\"odinger equation 
with some effective Hamiltonian, while the variable $R$ and the wave function $\chi(R)$ play an auxiliary   role and serve for the introduction of the quantum clock and, sometimes, of the  classical time.

Before the conclusion of this section we can mention that  some analogue of the classical time can be introduced even in the system with  one degree of freedom \cite{Sommerfeld, Rowe}. The idea is very simple.  Let us consider a particle with one spatial coordinate and a stable probability distribution for this coordinate. Naturally, the quantum state of such a particle is  an eigenstate of its Hamiltonian. Then, one can suppose that behind this probability distribution there is a classical motion which we can observe stroboscopically. That means that we can detect its position many times and obtain a probability distribution for this position. Classically
 this measured probability is inversely proportional to the velocity of the particle. Indeed, the higher is the velocity of a particle in some region of the space the less is the time that it spends there. 
 However in  quantum mechanics this probability is given by the squared modulus of its wave function. Thus,  we can write down the following equality \cite{Sommerfeld,Rowe}
\begin{equation}
\psi^*(x)\psi(x) = \frac{1}{|v(x)|T},
\label{classic}
\end{equation}
where $T$ is a some kind of normalising time scale, for example, a half period of the motion of the particle.  In this spirit, in  paper \cite{Rowe}, the probability distributions for the energy eigenstates of the hydrogen atom with a large principal quantum number $n$ were studied. It was shown that the distributions with the orbital quantum number $l$ having the maximal possible value $l=n-1$, being interpreted as in Eq. (\ref{classic}), describe the corresponding classical motion  of the electron on the circular orbit. At the same time, the state with $l=0$ cannot produce immediately a correct classical limit \cite{Rowe}.  To arrive to such a limit, which represents a classical radial motion of a particle
(i.e.  on a degenerate ellipse) one should apply a coarse-graining procedure based on the Riemann - Lebesgue theorem \cite{Rowe}. 
 There is another interesting example: the harmonic oscillator with a large value of the quantum number $n$. In this case, making a coarse-graining of the probability density one can again reproduce a classical motion of the oscillator \cite{Pauling}. We shall consider in detail this example and some other examples elsewhere \cite{we-future}. 

Usually, when one studies  the question of the classical-quantum correspondence, one looks for the situations where this correspondence is realised. However, it is reasonable to suppose that such situations are not always realised. Moreover, they  can be rather exceptional (for the respective discussion see e.g. \cite{Bar-Kam-quant} and  references therein).  
Here we would like to attract  attention to another phenomenon: a particular quantum-classical duality between the systems governed by different Hamiltonians. 

We can consider  a simple example. Let us suppose that we have a classical motion of the harmonic oscillator, governed by the law
\begin{equation}
x(t) = x_{0}\sin\omega t.
\label{harmon1}
\end{equation}
The velocity is 
\begin{equation}
\dot{x}(t)=\omega x_0\cos\omega t.
\label{velocity1}
\end{equation}
Using Eq. (\ref{classic}), 
we can suppose that associated with this classical motion is a stationary wave function 
\begin{equation}
\psi(x) = \frac{1}{\sqrt{\pi}(x_0^2-x^2)^{1/4}}e^{if(x)}\theta(x_0^2-x^2),
\label{function1}
\end{equation}
where $\theta$ is the Heaviside theta-function and $f$ is a real function.  Now, applying the energy conservation law and the stationary Schr\"odinger equation we can 
find the corresponding potential for the quantum problem:
\begin{eqnarray}
&&V(x) = \frac{m\omega^2x_0^2}{2}+\frac{\hbar^2}{2m}\left(\frac{1}{2(x_0^2-x^2)}+\frac{5x^2}{4(x_0^2-x^2)^2}+if''+if'\frac{x}{x_0^2-x^2}-f'^2\right),\nonumber \\
&&\ {\rm if}\ x^2 < x_0^2,
\label{potential1}
\end{eqnarray} 
Here, ``prime'' means the derivative with respect to $x$. To guarantee the reality of the potential and, hence, the hermiticity of the Hamiltonian, we must choose the phase function $f$ such that 
\begin{equation}
f' = C\sqrt{x_0^2-x^2},
\label{phase}
\end{equation}
where $C$ is a real constant.
Then the potential (\ref{potential1}) is equal to 
\begin{eqnarray}
&&V(x) = \frac{m\omega^2x_0^2}{2}+\frac{\hbar^2}{2m}\left(\frac{1}{2(x_0^2-x^2)}+\frac{5x^2}{4(x_0^2-x^2)^2}+C^2(x^2-x_0^2)\right),\nonumber \\
&&\ {\rm if}\ x^2 < x_0^2.
\label{potential2}
\end{eqnarray} 
Then for  $x^2>x_0^2$ we can treat the potential as an infinite since there the wave function is zero.  

Naturally, the example constructed above is rather artificial. 
We have elaborated it to  hint at the possibility of encountering a similar effect in cosmology. One 
can imagine  a situation where behind the visible classical evolution of the universe looms a quantum system, whose Hamiltonian  is quite   different 
from the classical Hamiltonian  governing this visible classical evolution.

\section{Concluding remarks}
In this short paper we have tried to compare two approaches to the introduction of time in  quantum cosmology. The first approach is based on the choice of a time-dependent gauge-fixing condition, which defines not only  the time parameter, but also a set of independent variables. Then, the  non-independent variables are expressed by means of the independent variables on solving the constraints and the non-vanishing  physical Hamiltonian arises. Subsequently, the independent variables are quantised and we obtained 
a physical Schr\"odinger equation. The solutions of this equations give us the probability distributions for the physical variables. 

The Born-Oppenheimer approach to the introduction of the time and to the description of the matter-gravity system in cosmology is based on the existence of two scales in the theory - one is the Planck scale, while the other is the scale characterising matter content of the Universe. Because of this one can suggest that the gravitational degrees of freedom 
have a slower evolution than the matter degrees of freedom. It allows one to omit some terms in the Wheeler-DeWitt equation and to introduce a time parameter due to the existence of the semiclassical Einstein (in our case - Friedmann) equations. This time parameter enters into the effective Schr\"odinger equation describing the evolution of the wave function of matter. It is important to note that in this case the effective physical Hamiltonian includes not only quantum operators, corresponding to the  matter variables, but also 
the quantum averages of these operators with respect to the wave function of matter, which we are looking for. We have applied these two methods to a very simple cosmological model, representing a flat Friedmann universe filled with the massless scalar field. In both cases as a time parameter we  have chosen the  cosmic time, corresponding to the choice of the lapse function $N=1$. 

In spite of the differences existing between these two approaches,  in this particular case they give similar results. 
It is not clear if  this correspondence will survive in more complicated models. Besides, it is not clear what is the counterpart of the second derivative $\frac{\partial^2\chi}{\partial\alpha^2}$ in the  approach based on the gauge-fixing procedure. We think that these questions deserve further study because 
the problem of self-consistent quantisation of systems consisting of gravity and matter are of great importance in modern theoretical physics.
The question of the classical-quantum correspondence is of great importance as well.

\section*{Acknowledgements}
We are grateful to A.O. Barvinsky for fruitful discussions. The work of A.K. was partially supported by  the RFBR grant No.17-02-00651.

\end{document}